\documentclass[structabstract]{aa} 
\usepackage{graphicx}
\usepackage{txfonts}
\usepackage{natbib}
\bibpunct{(}{)}{;}{a}{}{,}
\newcommand{\ros}{ROSAT}
\newcommand{\chan}{Chandra}
\newcommand{\xmm}{XMM-Newton}

\newcommand{\nh}{N_{\rm H}}

\def \msev{M7}
\def \magoneeig{\object{RX~J1856.5-3754}}
\def \magzersev{\object{RX~J0720.4-3125}}
\def \magonesix{\object{RX~J1605.3+3249}}
\def \magonethr{\object{RX~J1308.6+2127}}
\def \magtwoone{\object{RX~J2143.0+0654}}
\def \magzereig{\object{RX~J0806.4-4123}}
\def \magzerfou{\object{RX~J0420.0-5022}}
\def \ccoonee{\object{1E\,1207.4-5209}}

\def \jsixt{J1605}
\def \rrat{\object{RRAT~J1819-1458}}

\begin{document}
\title{\xmm\ reveals a candidate period for the spin of the ``Magnificent Seven'' neutron star \magonesix}
\author{A.~M.~Pires\inst{1}
    \and F.~Haberl\inst{2}
    \and V.~E.~Zavlin\inst{3}
    \and C.~Motch\inst{4}
    \and S.~Zane\inst{5}
    \and M.~M.~Hohle\inst{6}
    \fnmsep\thanks{Based on observations obtained with \xmm, an ESA science mission with instruments and contributions directly funded by ESA Member States and NASA (Target \magonesix, \textsf{\small obsid}~0671620101).}}
\offprints{A. M. Pires}
\institute{Leibniz-Institut f\"ur Astrophysik Potsdam (AIP), An der Sternwarte 16, 14482 Potsdam, Germany, 
    \email{apires@aip.de} 
    \and
    Max-Planck-Institut f\"ur extraterrestrische Physik, Giessenbachstra\ss e, 85748 Garching, Germany
    \and
    NASA Marshall Space Flight Center, Universities Space Research Association, VP62, Huntsville, AL 35812, USA
    \and 
    CNRS, Universit\'e de Strasbourg, Observatoire Astronomique, 11 rue de l'Universit\'e, F-67000 Strasbourg, France
    \and
    Mullard Space Science Laboratory, University College London, Holmbury St. Mary, Dorking, Surrey, RH5 6NT, UK
    \and
    Astrophysikalisches Institut und Universit\"ats-Sternwarte Jena, Schillerg\"asschen 2-3, 07745 Jena, Germany}
\date{Received ...; accepted ...}
\titlerunning{Discovery of pulsations from \magonesix}
\authorrunning{A.~M.~Pires et al.}
\abstract
{The group of seven thermally emitting isolated neutron stars (INSs) discovered by \ros\ and known as the ``Magnificent Seven'' (\msev) is unique among the various neutron star populations. Crustal heating by means of magnetic field decay and an evolutionary link with magnetars may explain why these objects rotate more slowly and have higher thermal luminosities and magnetic field intensities than standard rotation-powered pulsars of similar age.}
{The third brightest INS, \magonesix, is the only object amidst the seven still lacking a detected periodicity. The source spectrum, while purely thermal with no significant magnetospheric emission, is complex and displays both narrow and broad absorption features that can potentially be used to constrain the surface component of the magnetic field, as well as the mass-to-radius ratio of the neutron star.}
{We observed the source with the \xmm\ Observatory for 60\,ks aiming at unveiling the neutron star rotation rate and investigating its spectrum in detail. We confront our results with previous observations of the source and discuss its properties in the context of the \msev\ as a group and of the known population of Galactic INSs.}
{A periodic signal at $P=3.387864(16)$\,s, most likely the neutron star spin period, is detected at the $4\sigma$ confidence level. The amplitude of the modulation was found to be energy dependent and is more significantly detected when the timing search is restricted to photons with energy higher than $\sim0.5$\,keV. 
The coherent combination of the new data with a past \xmm\ EPIC-pn observation of the source constrains the pulsar spin-down rate at the $2\sigma$ confidence level, $\dot{\nu}\sim-1.39\times10^{-13}$\,Hz\,s$^{-1}$, implying a dipolar magnetic field of $B_{\rm dip}\sim7.4\times10^{13}$\,G. If confirmed, \magonesix\ would be the neutron star with the highest dipolar field amongst the \msev. The spectrum of the source shows evidence of a cool blackbody component, as well as for the presence of two broad absorption features. Furthermore, high-resolution spectroscopy with the RGS cameras confirms the presence of a narrow absorption feature at energy $\sim0.57$\,keV in the co-added spectrum of the source, also seen in other thermally emitting isolated neutron stars.}
{Phase-resolved spectroscopy, as well as a dedicated observing campaign aimed at determining a timing solution, will give invaluable constraints on the neutron star geometry and will allow one to confirm the high value of spin down, which would place the source closer to a magnetar than any other \msev\ INS.
}
\maketitle

\section{Introduction\label{sec_intro}}
Forty-five years after the discovery of PSR~B1919+21 \citep{hew68}, pulsars detected in radio surveys still dominate neutron star statistics \citep{man05}. Interestingly, it is at high energies that the neutron star phenomenology is revealed in all its complexity, with the discovery of peculiar classes of isolated neutron stars (INSs) not detected in the radio regime \citep[see][for an overview]{kas10,mer11,har13}. These include most of the magnetars -- anomalous X-ray pulsars (AXPs) and soft gamma repeaters (SGRs) -- the central compact objects (CCOs) in supernova remnants and the \ros-discovered thermally emitting INSs, also known as the ``Magnificent Seven'' (\msev).

In particular, the \msev\ is a remarkable group of INSs \citep[][for reviews]{hab07,kap08a,tur09}. As likely products of the nearby OB associations of the Gould Belt \citep[e.g.][]{wal01,pop03,mot03,mot05,pos08,mot09}, the sources are located within a few hundred parsecs of the Sun \citep{pos07,ker07a}. Their X-ray luminosity, dominated by thermal emission with no significant evidence of any magnetospheric activity, is believed to come directly from the neutron star surface. These properties make the \msev\ ideal targets for probing atmosphere models and cooling curves, and they can eventually be used to constrain the mass-to-radius ratio of the neutron star \citep[e.g.][]{ham11}. Unfortunately, the current lack of understanding of the surface composition, magnetic field, and temperature distributions have limited any definite conclusion \citep[][for an update on this issue]{kap11a}.

Broad features in absorption are observed in the X-ray spectra of most of the sources (the exceptions being the two softest members, \magoneeig\ and \magzerfou\footnote{We note that a spectral line in absorption at energy $\sim0.3$\,keV in the spectrum of the latter is disputed, probably as a result of uncertainties in the calibration of EPIC at the lowest energies; see \citealt{hab04b,kap11b} for details).}). The features are generally understood in terms of the neutron star magnetic field, although their interpretation is not unique. They can be related to cyclotron transitions of either protons ($B_{\rm cyc}\sim10^{13}-10^{14}$\,G) or electrons ($B_{\rm cyc}\sim10^{10}-10^{11}$\,G); while an alternative explanation would rest upon atomic transitions in the outermost layers of the neutron star \citep[e.g.][]{pot98,lai01a,med08}. 

Timing studies in X-rays \citep[see][and references therein]{kap11b} have shown that the \msev\ rotate slower ($P\sim3-10$\,s) and have higher magnetic field intensities ($B_{\rm dip}\sim{\rm few\ }\times10^{13}$\,G) than the bulk of the radio pulsar population.
Also at variance with pulsars detected at high energies, their X-ray luminosity is in excess of the spin-down power, suggesting that additional heating of the neutron star crust was at work by means of field decay \citep{agu08,pon09,vig13}. Therefore, these neutron stars may had experienced a different magneto-rotational evolution relative to standard pulsars, which consequently affected their cooling rates and detection as thermal X-ray sources \citep{kap09d,pop10}.

\magonesix\ (a.k.a RBS~1556, \citealt{schwope00}; \jsixt\ here for short), the third brightest INS \citep{mot99}, is the only source amidst the seven still lacking a detected periodicity. The source was observed by \xmm\ \citep{jan01} in ten different occasions\footnote{In four other observations, \jsixt\ was targeted for calibration purposes.} between January 2002 and February 2006, which amounted to nearly 260\,ks of observing time. Results of the first five observations were reported and discussed in detail by \citet{ker04}. The other five \xmm\ observations of the source, conducted in 2006, were severely affected by background flares, unfortunately reducing exposure times by more than $\sim80\%$ in the EPIC cameras. Likewise, the \chan\ X-ray Observatory \citep{wei02} targeted \jsixt\ in two occasions. The imaging ACIS observation, conducted in 2002, was also discussed by \citet{ker04}. The other \chan\ observation of the source, performed in 2007, aimed at high-resolution spectroscopy with the LETG instrument. 

Due to visibility constraints, the source could only recently be targeted again by the \xmm\ satellite.
We report here the results of a new observation, which was finally able to unveil a candidate spin for the neutron star. The paper is structured as follows: in Sect.~\ref{sec_datared} we describe the new \xmm\ observation and the data reduction. Data analysis and results are presented in Sect.~\ref{sec_analysis}. We discuss the implications of our findings in the context of the \msev\ as a group, and in the light of the properties of the observed population of Galactic neutron stars in Sect.~\ref{sec_discussion}. Our main results and conclusions are summarised in Sect.~\ref{sec_summary}.

\section{Observations and data reduction\label{sec_datared}}
The observation was carried out on 2012 March 6, for a total exposure time of 60.418\,ks. Table~\ref{tab_exposureinfoEPIC} contains information on the scientific exposures and instrumental setup of the RGS \citep{her01}, EPIC-pn \citep{str01} and EPIC-MOS \citep{tur01} detectors.
\subsection{EPIC data reduction\label{sec_epicdatared}}
The EPIC cameras were operated in full-frame mode with thin filters. Standard data reduction was performed with \textsf{\small SAS~13} (\textsf{\small xmmsas\_20130501\_1901-13.0.0}) using the latest calibration files.
\begin{table}[t]
\caption{Instrumental configuration and duration of the EPIC and RGS scientific exposures of the \xmm\ AO11 observation of \magonesix
\label{tab_exposureinfoEPIC}}
\centering
\begin{tabular}{l c c r}
\hline\hline
Instrument & Start Time & Mode & Duration \\
           & (UTC)      &      & (s)      \\
\hline
pn   & 2012-03-06T11:28:29 & Imaging      & 58,542 \\ 
MOS1 & 2012-03-06T11:06:14 & Imaging      & 47,892 \\ 
     & 2012-03-07T01:09:16 & Imaging      &  9,540 \\
MOS2 & 2012-03-06T11:06:14 & Imaging      & 47,899 \\
     & 2012-03-07T01:16:33 & Imaging      &  9,109 \\
RGS1 & 2012-03-06T11:05:30 & Spectroscopy & 60,418 \\
RGS2 & 2012-03-06T11:05:38 & Spectroscopy & 60,406 \\
\hline
\end{tabular}
\tablefoot{The EPIC cameras were operated in \textit{full frame} science mode and the thin filter was used. The two exposures in each MOS camera were merged for the analysis, unless otherwise noted. The RGS cameras were operated in \textit{high event rate with SES} spectroscopy mode for readout.}
\end{table}
We processed the MOS and pn raw event files using the EPIC meta tasks \textsf{\small emchain} and \textsf{\small epchain}, respectively, applying default corrections. 
We ensured that the pn event file was clean of unrecognised time jumps (i.e. those uncorrected by standard \textsf{\small SAS} processing). 
\begin{table*}[t]
\caption{Parameters of \magonesix, as extracted from the AO11 \xmm\ EPIC observations
\label{tab_sourceMLparam}}
\centering
\begin{tabular}{l r r r r}
\hline\hline
Parameter               & pn                                  & MOS1                              & MOS2                               & EPIC \\
\hline
Counts                  & $1.043(3)\times10^5$                & $2.906(21)\times10^4$             & $3.398(20)\times10^4$              & $1.679(5)\times10^5$ \\
Counts ($0.2-0.5$\,keV) & $6.023(24)\times10^4$               & $1.779(16)\times10^4$             & $2.051(16)\times10^4$              & $9.89(4)\times10^4$ \\
Counts ($0.5-1.0$\,keV) & $4.145(21)\times10^4$               & $1.033(13)\times10^3$             & $1.233(12)\times10^4$              & $6.430(28)\times10^4$ \\
Counts ($1.0-2.0$\,keV) & $2.61(6)\times10^3$                 & $9.3(4)\times10^2$                & $1.14(4)\times10^3$                & $4.70(8)\times10^3$ \\
Counts ($2.0-4.5$\,keV) & $4\pm4$                             & $6\pm4$                           & $1.0\pm2.0$                        & $12(6)$ \\
Counts ($4.5-12$\,keV)  & $0.0\pm2.0$                         & $0.0\pm2.0$                       & $0.0\pm2.0$                        & $0(3)$ \\
Detection likelihood    & $589,256$                           & $112,983$                         & $181,949$                          & $881,737$ \\
Rate (s$^{-1}$)         & $3.018(9)$                          & $0.695(5)$                        & $0.677(4)$                         & $4.413(12)$ \\
RA                      & $16$\ \ $05$\ \ $18.366\pm0.029''$  & $16$\ \ $05$\ \ $18.29\pm0.06''$  &  $16$\ \ $05$\ \ $18.30\pm0.05''$  &  $16$\ \ $05$\ \ $18.5\pm0.3''$\,$^{(\dagger)}$ \\
DEC                     & $+32$\ \ $49$\ \ $18.673\pm0.029''$ & $+32$\ \ $49$\ \ $18.02\pm0.06''$ &  $+32$\ \ $49$\ \ $18.48\pm0.05''$ &  $+32$\ \ $49$\ \ $19.2\pm0.3''$\,$^{(\dagger)}$ \\
$l$ (degrees)           & $52.881$                            & $52.881$                          & $52.881$                           & $52.881$ \\
$b$ (degrees)           & $+47.993$                           & $+47.993$                         & $+47.993$                          & $+47.993$ \\
HR$_1$                  & $-0.185\pm0.003$                    & $-0.264\pm0.007$                  & $-0.249\pm0.006$                   & $-0.2118(27)$ \\
HR$_2$                  & $-0.8817\pm0.0024$                  & $-0.834\pm0.006$                  & $-0.864\pm0.021$                   & $-0.8639(21)$ \\
HR$_3$                  & $-0.997\pm0.003$                    & $-0.987\pm0.0.008$                & $-0.995\pm0.027$                   & $-0.9950(27)$ \\   
\hline
\end{tabular}
\tablefoot{Counts and rates are given in the total \xmm\ energy band ($0.2-12$\,keV), unless otherwise specified. $^{(\dagger)}$Corrected with the \textsf{SAS} task \textsf{eposcorr}, based on a number of 46 X-ray sources cross-correlated with the GSC~2.3.2 catalogue (see text).}
\end{table*}

Background flares were registered during the beginning and towards the end of the EPIC observation. The effective observing times, after filtering out periods with background levels higher than those recommended by standard \xmm\ data reduction, are 40\,ks for pn and 51\,ks for MOS.
For the analysis, we filtered the event lists to exclude bad CCD pixels and columns, as well as to retain the pre-defined photon patterns with the highest quality energy calibration. Unless otherwise noted, single and double events were selected for pn (pattern $\le4$) and single, double, triple, and quadruple for MOS (pattern $\le12$). We defined the source centroid and optimal extraction region for each EPIC camera with the task \textsf{\small eregionanalyse} in the energy band $0.2-1.2$\,keV.
Background circular regions of size $60''$ to $100''$ were defined off-source, on the same CCD as the target. 

The detected source count rates, hardness ratios, and other parameters based on a maximum likelihood fitting (as determined with the \textsf{\small SAS} task \textsf{\small emldetect} on the EPIC images in each camera and energy band) are listed in Table~\ref{tab_sourceMLparam}. 
We used the task \textsf{\small eposcorr} to refine the astrometry, by cross-correlating the list of EPIC X-ray sources in the field-of-view with those of optical (GSC~2.3.2, \citealt{las08}) objects lying within $15'$ from \jsixt. We found offsets of $-1.6''\pm0.3''$ in right ascension and $-0.5''\pm0.3''$ in declination, based on a number of forty-six X-ray/optical matches. 
The corrected EPIC source position (Table~\ref{tab_sourceMLparam}), updated accordingly, was found consistent within errors with previous determinations \citep{kap03b}, taking into account the known proper motion of the source (derived from the more accurate observations of the optical counterpart; \citealt{mot05,zan06b}).
The statistics for the EPIC lightcurves (corrected for bad pixels, deadtime, exposure, as well as background counts, and binned into 600\,s intervals) show the $3\sigma$ upper limits for the r.m.s. fractional variation of 0.018, 0.04 and 0.05 for pn, MOS1, and MOS2, respectively. 
On the basis of the same lightcurves, the reduced $\chi^2_\nu$ assuming a constant flux is 1.17 (pn, 78 d.o.f.), 1.14, and 0.90 (MOS1 and MOS2; 90 and 89 d.o.f.), corresponding to null-hypothesis probabilities of 14\%, 17\%, and 74\%. 
\subsection{RGS data reduction\label{sec_rgsdatared}}
We processed the RGS data using the \textsf{\small SAS} routine \textsf{\small rgsproc}. Following standard procedure, we identified times of low background activity from the count rate on CCD~9, the closest to the optical axis, and applied a count rate threshold of 0.1\,s$^{-1}$ to filter the good-time intervals. The final exposure times and net count rates are $\sim51$\,ks and $0.1053(16)$\,s$^{-1}$, and $\sim52$\,ks and $0.08852(15)$\,s$^{-1}$, in RGS1 and RGS2, respectively, in the $0.35-1.0$\,keV energy band. 
Due to electronic problems\footnote{\texttt{http://xmm2.esac.esa.int/docs/documents/\\CAL-TN-0030.pdf}}, one CCD chip of each of the RGS detectors failed early in the mission; these affect the spectral coverage between $11$\,\AA\ and $14$\,\AA\ and between $20$\,\AA\ and $24$\,\AA, respectively, in RGS1 and RGS2. 

\section{Analysis and results\label{sec_analysis}}
\subsection{Unveiling the neutron star spin period\label{sec_timinganalysis}}
Searches for a periodic signal in previous observations of the source revealed no statistically significant candidate for the neutron star spin; the derived upper limit on the pulsed fraction was $p_{\rm f}\lesssim5\%$ at the 95\% confidence level, in the frequency range of $0.01$\,Hz to $800$\,Hz \citep{ker04}. It is worth noting that the EPIC pn observations on which these previous studies were based were performed in timing mode, which requires a larger photon extraction area, consequently resulting in a higher background and lower signal-to-noise ratio in the period search.
\begin{figure}
\begin{center}
\includegraphics[width=0.495\textwidth]{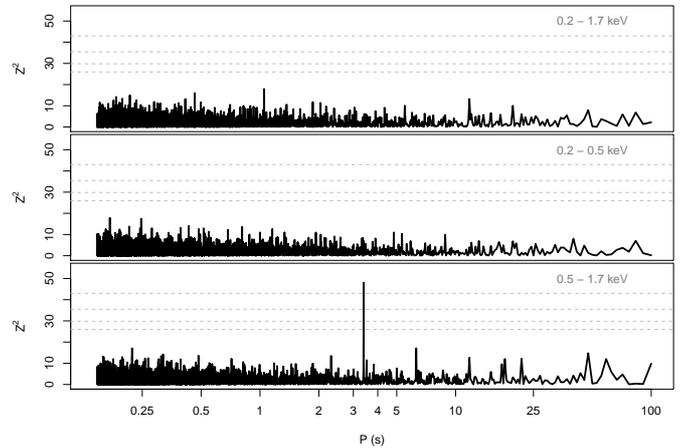}
\end{center}
\caption{Results of the $Z^2_1$ search (pn data, $P=0.1468-100$\,s). The frequency range is $\Delta\nu\sim6.8$\,Hz, the energy bands are $0.2-1.7$\,keV (top), $0.2-0.5$\,keV (centre) and $0.5-1.7$\,keV (bottom). The size of the extraction region is $9''$ amounting to $\sim119,000$, $\sim82,500$ and $\sim36,800$ counts, respectively, in those bands. Dashed horizontal lines show confidence levels of $1\sigma$ to $4\sigma$ for the detection of a periodic signal, given the frequency range, duration of the observation and the number of source photons in the search. A periodic signal at $P_\star\sim3.39$\,s is detected at $>4\sigma$ when the energy band is restricted to the harder source photons.}\label{fig_z2result}
\end{figure}

For the timing analysis, we considered events with pattern 12 or lower, and tested different energy bands and radii of the source extraction region. To achieve the maximum sensitivity of our data, we used the event lists of the three EPIC cameras unfiltered for background flares (Sect.~\ref{sec_epicdatared}), after checking that the soft energy band is not severely affected by them\footnote{We note that results of the timing analysis are unchanged when using the filtered event list sets (see main text).}. To avoid aliasing artefacts in the timing power spectra, only the first (and longest) exposures in each of the MOS cameras were considered, since the merged event files have a $\sim5$\,ks gap in between exposures, after the event of a background flare (Table~\ref{tab_exposureinfoEPIC}). The times-of-arrival of the pn/MOS photons were converted from the local satellite to the solar system barycentric frame using the \textsf{\small SAS} task \textsf{\small barycen} and the source coordinates in each camera (Table~\ref{tab_sourceMLparam}). A $Z^2_n$ (Rayleigh) test \citep{buc83} was applied to search for pulsations. The adopted step in frequency was 2\,$\mu$Hz (or an oversampling factor of 10), warranting that a peak corresponding to a periodic signal is not missed.

In full-frame mode, pn and MOS provide a time resolution of 73.4\,ms and 2.6\,s, respectively, with negligible deadtime. Therefore, searches conducted over photons from the three EPIC cameras are restricted to periodicities longer than $P>5.2$\,s. We found no significant pulsations, with pulsed fraction higher than 1.6\% ($3\sigma$), in this period range (total of $\sim240,600$ photons in the energy band $0.2-1.5$\,keV; the number of independent trials is $\sim10^4$). To look for higher frequency pulsations, with periods as short as $P\sim0.15$\,s, we restricted the analysis to the pn camera and performed more extensive searches.

Very interestingly, we found that a periodic signal at frequency $\nu\equiv\nu_\star=0.2951712(14)$\,Hz is revealed when the search is restricted to photons with energy higher than $\sim0.5$\,keV. The same frequency, however with a less significant $Z^2_1$ power, is also the highest when the test is performed over the background-filtered event lists or with different photon patterns and radii of the source extraction region (we tested photon patterns 0, 4 and 12 and radii between $5''$ and $40''$). We ensured that no high power is detected when the search is performed over photons extracted from several background regions (in several energy bands and with roughly the same total number of counts as collected for \jsixt) hence excluding the possibility that the signal is associated with unknown instrumental effects.

We refined the search around the found periodicity in order to find the best choice of parameters that maximises the power of the $Z^2_1$ test. We found that the peak at period $P\equiv P_\star=3.387864(16)$\,s has a maximum power of $Z^2_{\rm max}\sim50$ in the energy band of $0.5-1.69$\,keV and for an extraction radius of $9''$, which amounts to $N_{\rm ph}\sim3.69\times10^4$ photons in the search (Fig.~\ref{fig_z2result}, bottom); the $Z^2_{\rm max}$ power corresponds to a detection at $4.3\sigma$ for this choice of parameters ($3.25\times10^5$ independent trials in the $\nu=0.01-6.81$\,Hz frequency range). By contrast, no frequency with pulsed fraction higher than $p_{\rm f}=3\%$ ($3\sigma$) shows a significant power in the $Z^2_1$ test in either the energy band where the bulk of the source photons are emitted, $0.2-1.7$\,keV, or in the very soft energy range of $0.2$\,keV to $0.5$\,keV (total of $\sim119,000$ and $\sim82,500$ pn events, respectively; see top and centre plots of Fig.~\ref{fig_z2result}); this suggests that most of the source photons show very low-amplitude modulation and easily smear out the significance of the signal detected at harder energies. It is nonetheless interesting to note that the found periodicity is within the narrow range of measured spin periods of the other \msev\ (e.g. \citealt{hab07}), which we discuss further in Sect.~\ref{sec_discP}. 
\begin{figure}
\begin{center}
\includegraphics*[width=0.475\textwidth]{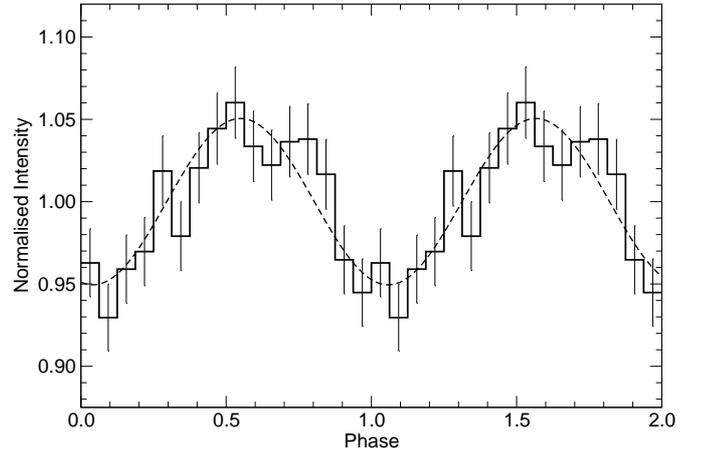}
\end{center}
\caption{Background-subtracted light curve for pn, folded at the spin period $P\sim3.39$\,s. Two cycles are shown for clarity. The energy band is $0.5-1.69$\,keV and the size of the extraction region is $9''$. A best-fit sinusoidal curve to the folded light curve is superposed in dashed line.}\label{fig_folded}
\end{figure}

The pn lightcurve in the energy band $0.5-1.69$\,keV, folded at the found periodicity and corrected for background counts and other effects, can be seen in Fig.~\ref{fig_folded}. The pulsed fraction computed from the best sinusoidal fit is $p_{\rm f}=5.1(7)\%$, in agreement with the results from the $Z^2_1$ analysis with optimised search parameters, $p_{\rm f}=(2Z^2_{\rm max}/N_{\rm ph})^{1/2}\times 100\% \sim 5\%$. The inclusion of higher harmonics $n\ge2$ in the $Z^2_n$ test was found to be statistically insignificant, as evident by the sinusoidal pulse profile of the modulation. 

\subsection{Pulsar spin down\label{sec_spindown}}
We checked if the candidate period could had been detected in previous \xmm\ observations of \jsixt. 
The only past pn observation of the source suitable for timing analysis\footnote{Other archival EPIC-pn observations of the source were either conducted in timing mode, operated with the thick filter, or severely affected by background flares.} is the one performed in 2003 (\textsf{\small obsid} 0157360401), with an exposure time of $\sim33$\,ks. We processed and analysed this observation in the same way as described in Sect.~\ref{sec_epicdatared} and \ref{sec_timinganalysis}. By adopting the best energy band and parameters as found for the AO11 data, we extracted a number of $\sim19,800$ source photons for the timing analysis, which were barycentre corrected accordingly.

Accounting for the time span of $T_{\rm span}\sim2.88\times10^8$\,s between the two datasets and a $3\sigma$ uncertainty on $\nu_\star$, we searched for significant peaks allowing for a maximum braking corresponding to that of a pulsar with $B_{\rm dip}=10^{14}$\,G (assuming the usual pulsar spin-down formula of magnetic dipole braking in vacuum, see e.g. Sect.~\ref{sec_discussion}).
We found no significant pulsation in the frequency range of the search. The $4\sigma$ upper limit on the pulsed fraction, $p_{\rm f}=4.6\%$, shows that the 2003 observation is just as the limiting sensitivity to detect the shallow modulation found in the longer dataset.
\begin{table*}[t]
\caption{\small Summary of \xmm\ observations of \magonesix\ used in the spectral analysis \label{tab_pastxmmobs}}
\centering
\begin{tabular}{c c c c c c c r c r c l c l}
\hline\hline
Ref. & OBSID & MJD & \multicolumn{3}{c}{Science Mode} & & \multicolumn{3}{c}{Net Exposure (ks)} & & \multicolumn{3}{c}{Observed flux (10$^{-12}$\,erg\,s$^{-1}$\,cm$^{-2}$)}\\
\cline{4-6}\cline{8-10}\cline{12-14}
 &      & (days) & MOS & pn & RGS & & \multicolumn{1}{c}{MOS} & \multicolumn{1}{c}{pn} & \multicolumn{1}{c}{RGS} & & \multicolumn{1}{c}{MOS} & \multicolumn{1}{c}{pn} & \multicolumn{1}{c}{RGS} \\
\hline
A & 0073140301  & 52284.146834 & FF & TI & SES & & 25.156 & --     & 17.855 & & $6.06(8)$     & --          & $4.50(24)$ \\
B & 0073140201  & 52290.127905 & FF & TI & SES & & 27.949 & --     & 27.254 & & $6.08(8)$     & --          & $4.51(23)$ \\
C & 0073140501  & 52294.123831 & FF & TI & SES & & 22.102 & --     & 21.742 & & $6.09(10)$    & --          & $4.4(9)$   \\
D & 0157360401  & 52657.139334 & FF & LW & SES & & 30.094 & 27.797 & 28.898 & & $6.66(9)$     & $6.636(25)$ & $4.66(28)$ \\ 
E & 0157360601  & 52696.990896 & FF & LW & SES & &  7.250 & --     &  7.742 & & $6.50(16)$    & --          & $4.6(3)$   \\
AO11 & 0671620101  & 55992.811793 & FF & FF & SES & & 51.407 & 39.875 & 52.125 & & $6.12(9)$  & $6.610(24)$ & $4.12(15)$ \\
\hline
\end{tabular}
\tablefoot{The net exposure time per camera (averages are considered for the two MOS and RGS detectors) is filtered for periods of high background activity. All the EPIC observations included in the analysis made use of the thin filter. The instrument science modes are full-frame (FF), timing (TI), large window (LW), and spectroscopy (SES). The observed flux is in energy band $0.2-12$\,keV (EPIC) and $0.35-2.5$\,keV (RGS).}
\end{table*}

Regardless of the lower sensitivity of the 2003 observation, the data can still be used in combination with the AO11 observation in a coherent two-dimensional $Z^2_1(\nu,\dot{\nu})$ search, i.e. allowing the $Z^2$ test to account for the neutron star spin down \citep[see e.g.][for details]{pav99}. In this case, the ephemeris parameters $\nu,\dot{\nu}$ that determine the phase $\phi_j$ of each photon time-of-arrival, 

\begin{displaymath}
\phi_j=\nu(t_j-t_0) + \dot{\nu}\frac{(t_j-t_0)^2}{2}\qquad j=1,\ldots,N_{\rm ph}
\end{displaymath}

\noindent are estimated as the values that give the highest power of the $Z^2_1$ statistics (similarly as before, $t_0$ is the event time-of-arrival counted from an epoch of zero phase and $N_{\rm ph}=56,645$ is the total number of photons in the search when joining the two datasets).
\begin{figure}
\begin{center}
\includegraphics*[width=0.495\textwidth]{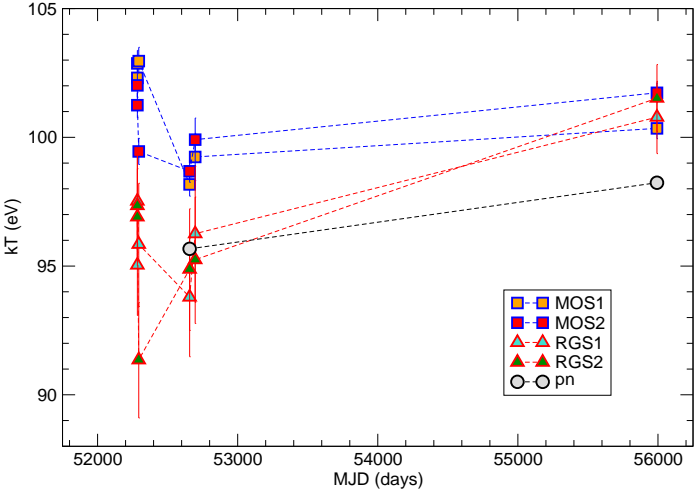}
\end{center}
\caption{Best-fit blackbody temperature of \magonesix\ in each observation and \xmm\ detector of Table~\ref{tab_pastxmmobs} (see caption). Errors are $1\sigma$.\label{fig_kTevol}}
\end{figure}

Again allowing for a maximum braking that corresponds to a field of $B_{\rm dip}=10^{14}$\,G, $|\dot{\nu}_{\rm max}|\equiv2.5\times10^{-13}$\,Hz\,s$^{-1}$, we searched the relevant parameter space in steps of $10^{-9}$\,Hz and $3.6\times10^{-18}$\,Hz\,s$^{-1}$; the number of independent trials is $\mathcal{N}=(\nu_{\rm max}-\nu_{\rm min})|\dot{\nu}_{\rm max}-\dot{\nu}_{\rm min}|T_{\rm span}^3/2\sim4.6\times10^8$. We found that the highest peak, $Z^2\sim46.3$ occurs at $(\nu,\dot{\nu})\sim(0.295211214{\rm\,Hz},-1.386075\times10^{-13}{\rm\,Hz\,s}^{-1})$, and corresponds to a fractional amplitude of $p_{\rm f}=4\%$, overall consistent with our results in the individual datasets. The significance of the detection is $2\sigma$; no other $(\nu,\dot{\nu})$ pair was found above the $1\sigma$ confidence level.

\subsection{Spectral analysis\label{sec_specanalysis}}
The spectral energy distribution of \jsixt\ is known to deviate from a pure blackbody, due to the presence of a broad line in absorption at energy $\sim0.45$\,keV \citep{ker04}. Similar features are also reported in the spectra of other thermally emitting INSs \citep{san02,hab03,zan04,mor05,hab07,schwope07,lau07,zan11,pir12}. 
Evidence for the presence of a narrow absorption spectral feature at energy $\sim0.57$\,keV, possibly related to either interstellar or circumstellar highly-ionised oxygen, was found in the co-added RGS spectra of the source \citep{ker04,hoh12a}.
\subsubsection{EPIC data}
The analysis of the EPIC data is based on source and background spectra extracted from regions as described in Sect.~\ref{sec_epicdatared}, together with the respective response matrices and ancillary files created for each of the EPIC cameras.
We also included in the spectral analysis the past pn and MOS exposures of the source conducted with the thin filter and not severely affected by background flares (Table~\ref{tab_pastxmmobs}; the archival observations are listed `A' to `E' as reference for the text). The available data amount to a total of 14 spectra and $4.1\times10^5$ counts in energy band $0.2-1.5$\,keV (of which $\sim2.5\%$ can be ascribed to background). We reprocessed and reduced all observations in consistency with the AO11 dataset.

Due to the brightness of the target and the use of the thin filter, approximately $\sim3.6\%$ and $1.5\%$ of the pn photons in the AO11 and `D' observations, respectively, were affected by pile-up. To minimise spectral distortion, a correction in the redistribution matrix files of the pn camera (calculated from the frequency and spectrum of the incoming photons) was applied\footnote{As of \textsf{\small SAS~13}, the task \textsf{\small rmfgen} includes an option to correct for the flux loss and energy distortion caused by the pile-up of photons within a single frame (only available for pn imaging).}. 
For all spectra, energy channels within $0.3$\,keV and $1.35$\,keV were rebinned according to a minimum number of 30 counts per spectral bin and we took care not to oversample the instrument energy resolution at a given bin by more than a factor of three, which is especially important at soft energies.
\begin{table*}[t]
\caption{Results of spectral fitting
\label{tab_resultspec}}
\centering
\begin{tabular}{c r r r r r r r r c r}
\hline\hline
 & $\chi^2_{\nu}$\,(d.o.f.) & \multicolumn{1}{c}{$\nh$} & \multicolumn{6}{c}{Parameters} & $EW$ & \multicolumn{1}{c}{$F_{\rm X}$} \\
\cline{4-9}
 & & & $kT^\infty_{\rm cool}$ & $kT^\infty_{\rm hot}$ & $\epsilon_1$ & $\sigma_1$ & $\epsilon_2$ & $\sigma_2$ & & \\
 & & ($\times10^{20}$) & (eV) & (eV) & (keV) & (keV) & (keV) & (keV) & (eV) & ($\times10^{-12}$) \\
\hline
\multicolumn{11}{l}{AO11 (pn)}\\
(1) & 17\,(25)  & 0                      & $-$                  & $100.68(25)$ & $-$                      & $-$                         & $-$                       & $-$                      & $-$     & $6.333(29)$ \\
(2) & 6.9\,(22) & 0                      & $-$                  & $98.2(8)$    & $0.462(28)$              & $0.101(16)$                 & $-$                       & $-$                      & $46$    & $6.8(3)$ \\
(3) & 2.4\,(19) & 0                      & $-$                  & $100.0(2.0)$ & $0.41(7)$                & $0.14(3)$                   & $0.8(5)$                  & $0.099(15)$              & $92,59$ & $7.1(1.0)$ \\
(4) & 3.4\,(23) & 0                      & $42(5)$              & $105.8(7)$   & $-$                      & $-$                         & $-$                       & $-$                      & $-$     & $8(7)$ \\ 
(5) & 1.9\,(20) & $1.60_{-0.27}^{+0.26}$ & $76.9(9)$            & $124.8(2.2)$ & $0.39_{-0.06}^{+0.10}$   & $0.0961_{-0.010}^{+0.0018}$ & $-$                       & $-$                      & $56$    & $9.85(22)$ \\
(6) & 1.7\,(18) & $^\star2.4$            & $62.3_{-2.1}^{+1.9}$ & $110.2(1.1)$ & $0.394_{-0.021}^{+0.12}$ & $0.087_{-0.015}^{+0.003}$   & $0.868_{-0.025}^{+0.018}$ & $0.028_{-0.028}^{+0.03}$ & $43,13$ & $12.1(5)$ \\
\hline
\multicolumn{11}{l}{AO11 (EPIC)}\\
(1) & 9.0\,(92)   & $0$         & $-$                  & $101.87(22)$          & $-$                    & $-$                         & $-$                       & $-$                       & $-$      & $5.98(8)$ \\
(2) & 5.0\,(89)   & $0$         & $-$                  & $99.3(6)$             & $0.473(16)$            & $0.088(10)$                 & $-$                       & $-$                       & $40$     & $6.5(4)$ \\
(3) & 2.1\,(86)   & $0$         & $-$                  & $100.2(1.6)$          & $0.4(4)$               & $0.138(24)$                 & $0.8(3)$                  & $0.099(10)$               & $99,64$  & $6.8(1.4)$ \\
(4) & 2.7\,(90)   & $0.1(1.0)$  & $38.6(2.4)$          & $106.6(9)$            & $-$                    & $-$                         & $-$                       & $-$                       & $-$      & $9(11)$ \\ 
(5) & 1.6\,(87)   & $2.07(23)$  & $72.4(7)$            & $122.3_{-1.2}^{+1.3}$ & $0.39_{-0.19}^{+0.22}$ & $0.0914_{-0.008}^{+0.0011}$ & $-$                       & $-$                       & $63$     & $10.1(8)$ \\
(6) & 1.5\,(85)   & $^\star2.4$ & $59.2_{-1.7}^{+1.5}$ & $110.2(7)$            & $0.400(25)$            & $0.077_{-0.014}^{+0.005}$   & $0.855_{-0.016}^{+0.013}$ & $0.054_{-0.016}^{+0.019}$ & $42,18$  & $11.9_{-0.8}^{+0.5}$ \\
\hline
\multicolumn{11}{l}{combined (EPIC)}\\
(1) & 70\,(33)  & $0$                      & $-$         & $101.19(12)$ & $-$                       & $-$                       & $-$                 & $-$        & $-$       & $5.791(15)$ \\
(2) & 41\,(30)  & $0$                      & $-$         & $99.2(6)$    & $0.4(7)$                  & $0.169(15)$               & $-$                 & $-$        & $67$      & $6.3(3)$ \\
(3) & 14\,(27)  & $0$                      & $-$         & $99.7(1.1)$  & $0.4(5)$                  & $0.146(16)$               & $0.79(19)$          & $0.101(6)$ & $106,70$  & $6.7(5)$ \\
(4) & 11\,(30)  & $0$                      & $34.7(1.1)$ & $106.15(23)$ & $-$                       & $-$                       & $-$                 & $-$        & $-$       & $10.4(1.7)$ \\ 
(5) & 2.0\,(28) & $0.014_{-0.014}^{+0.13}$ & $77.6(4)$   & $128.8(1.0)$ & $0.39_{-0.18}^{+0.19}$    & $0.0875(9)$               & $-$                 & $-$        & $56$      & $7.58_{-0.10}^{+0.11}$ \\
(6) & 1.5\,(25) & $0$                      & $44.8(3)$   & $109.6(5)$   & $0.402_{-0.04}^{+0.022}$  & $0.071_{-0.013}^{+0.008}$ & $0.8_{-0.8}^{+0.9}$ & $0.139(3)$ & $21,46$   & $8.67(22)$ \\
\hline
\multicolumn{11}{l}{Best fit (6) per OBSID in Table~\ref{tab_pastxmmobs}}\\
A & 1.3\,(58) & $^\star2.4$         & $35_{-4}^{+5}$       & $106.6_{-1.0}^{+1.2}$ & $0.442_{-0.03}^{+0.018}$  & $0.025_{-0.025}^{+0.018}$ & $0.81_{-0.06}^{+0.03}$ & $0.103_{-0.024}^{+0.022}$ & $6,30$  & $18.7_{-0.5}^{+17}$ \\
B & 1.0\,(57) & $1.5_{-1.2}^{+1.3}$ & $37_{-5}^{+7}$       & $106.6_{-1.1}^{+1.2}$ & $0.446_{-0.016}^{+0.012}$ & $0.028(12)$               & $0.8(6)$               & $0.094(11)$               & $15,41$ & $13_{-9}^{+19}$ \\  
C & 1.3\,(56) & $1.9_{-1.3}^{+1.2}$ & $45_{-6}^{+5}$       & $110.3_{-1.5}^{+1.7}$ & $0.445_{-0.019}^{+0.010}$ & $0.021_{-0.022}^{+0.018}$ & $0.86_{-0.08}^{+0.04}$ & $0.11_{-0.03}^{+0.04}$    & $98,42$ & $12_{-3}^{+9}$ \\
D & 1.1\,(85) & $^\star2.4$         & $39_{-4}^{+5}$       & $105.0_{-1.2}^{+1.3}$ & $0.36_{-0.22}^{+0.7}$     & $0.090_{-0.023}^{+0.005}$ & $0.7(1.0)$             & $0.167(4)$                & $28,56$ & $19_{-6}^{+10}$\\
E & 1.2\,(47) & $^\star2.4$         & $30.3_{-2.2}^{+2.4}$ & $110.5_{-2.7}^{+2.9}$ & $0.6(1.0)$                & $0.04(4)$                 & $0.77_{-0.29}^{+0.7}$  & $0.173_{-0.05}^{+0.014}$  & $2,66$  & $27_{-11}^{+24}$\\
\hline
\end{tabular}
\tablefoot{Parameters marked with a star are held fixed during fitting. The column density is in cm$^{-2}$. The unabsorbed flux is in erg\,s$^{-1}$\,cm$^{-2}$ in energy band $0.2-12$\,keV. Uncertainties on parameters for fits with a high value of reduced chi-square ($\chi^2_\nu>2$) shall be regarded as only indicative of the real confidence levels. Models: \textsf{(1)~bbody}, \textsf{(2)~bbody-gauss}, \textsf{(3)~bbody-2*gauss}, \textsf{(4)~2*bbody}, \textsf{(5)~2*bbody-gauss}, \textsf{(6)~2*bbody-2*gauss}.}
\end{table*}

To fit the spectra, we used \textsf{\small XSPEC~12.7.1} \citep{arn96}. The photoelectric absorption model and elemental abundances of \citet[][\textsf{\small tbabs} in \textsf{\small XSPEC}]{wil00} were adopted to account for the interstellar absorption. We note that the choice of abundance table and cross-section model in \textsf{\small XSPEC} does not impact the results of spectral fitting significantly, due to the low absorption towards \jsixt. We fitted the data of each camera and observation individually; additionally, to more tightly constrain the spectral parameters, we performed simultaneous fits, where we allowed for a renormalisation factor to account for cross-calibration uncertainties between the detectors and possible time variability of the spectral parameters. 
\begin{figure*}
\begin{center}
\includegraphics*[width=0.49\textwidth]{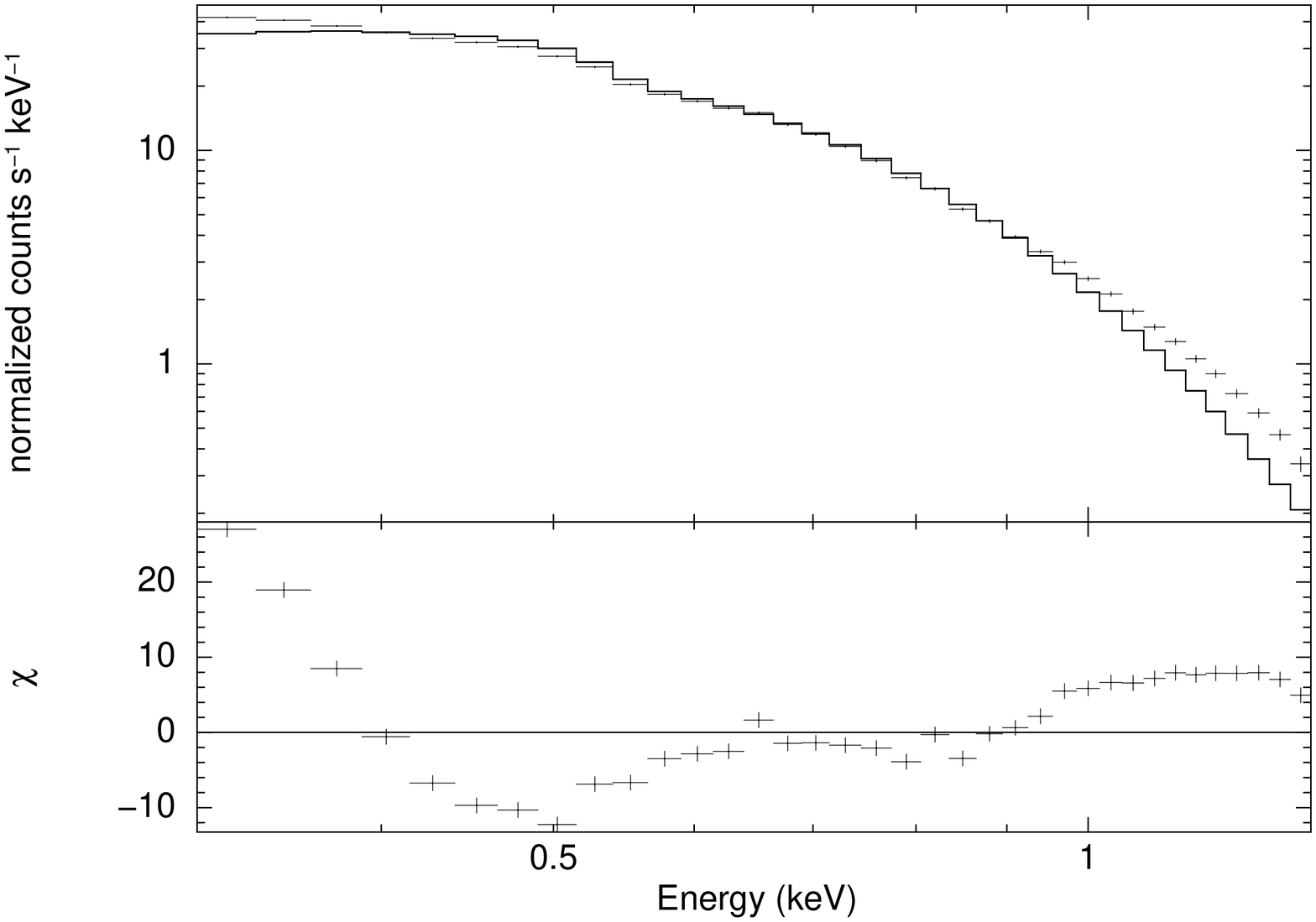}\hspace{0.25cm}
\includegraphics*[width=0.49\textwidth]{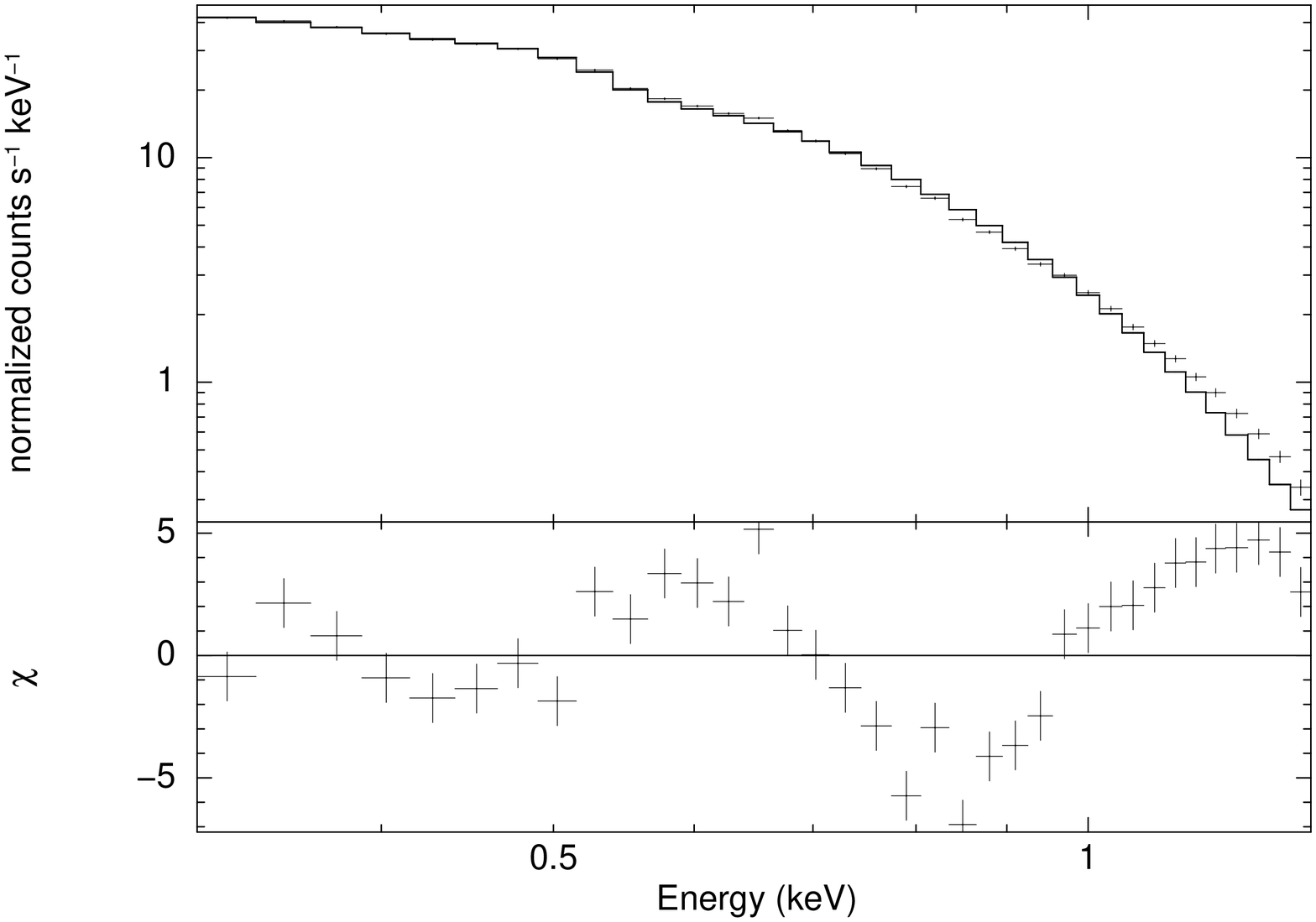}\\
\includegraphics*[width=0.49\textwidth]{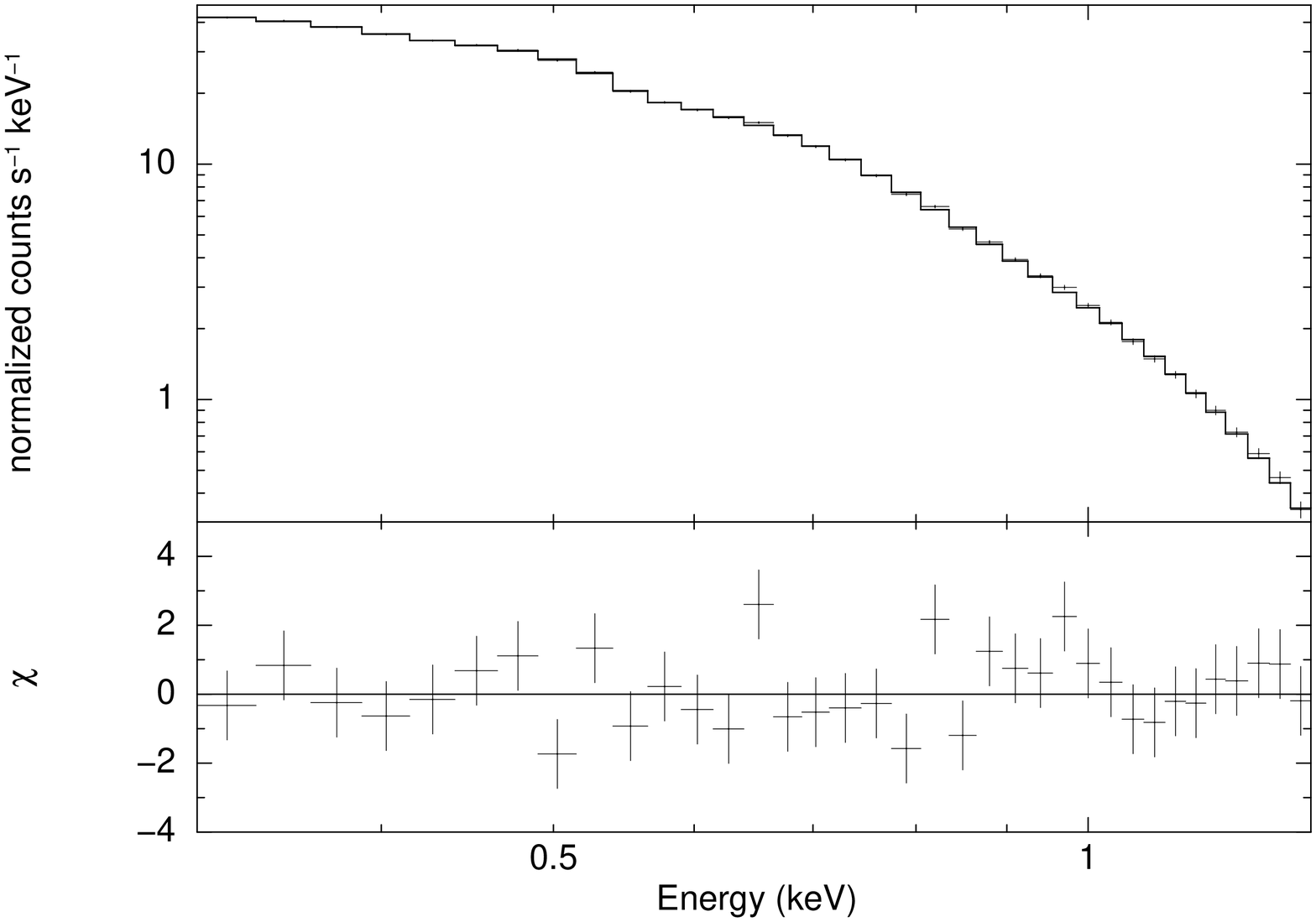}\hspace{0.25cm}
\includegraphics*[width=0.49\textwidth]{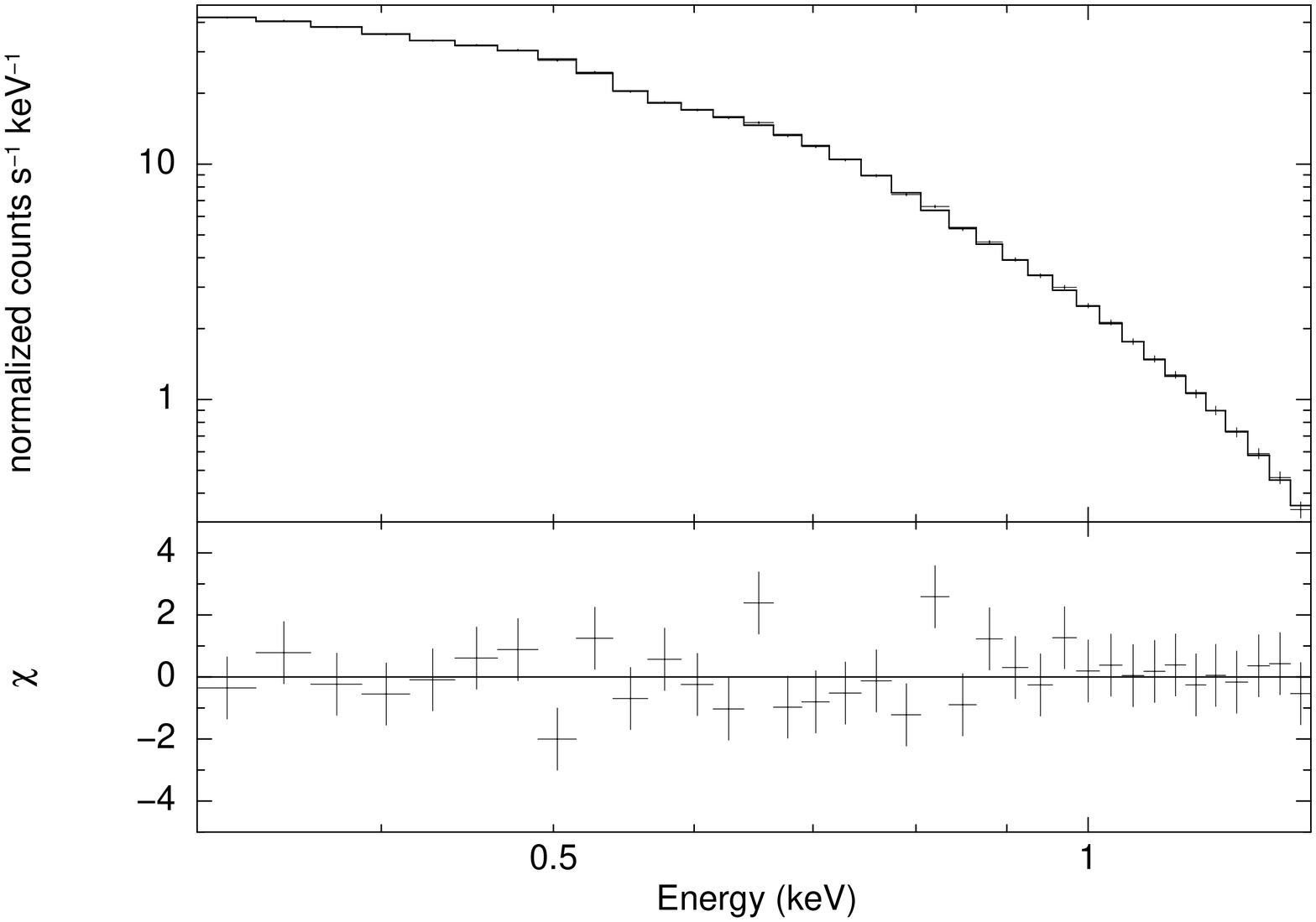}
\end{center}
\caption{Results of spectral fitting of source \jsixt\ (EPIC). We show the combined EPIC spectra and folded best-fit models, with residuals. \textit{Top left}: Simple blackbody model. \textit{Top right}: Double blackbody model. \textit{Bottom left}: Double blackbody model with one Gaussian absorption line. \textit{Bottom right}: Double blackbody model with two Gaussian absorption lines.\label{fig_spec}}
\end{figure*}

The observed flux of \jsixt\ is stable between the two pn observations, and consistent with a constant value (Table~\ref{tab_pastxmmobs}). Similarly, there is no evidence for flux variability in the RGS1/2 data (Sect.~\ref{sec_rgsspec}). 
On the other hand, significant flux and temperature variations are seen between the MOS1/2 exposures, as well as between the two detectors for a given epoch (see Fig.~\ref{fig_kTevol}, where we plot the best-fit $kT_\infty$ of a simple blackbody model for each observation and \xmm\ detector as a function of time).
The MOS CCDs are known to suffer from redistribution changes and contamination with time, making the instrument unsuitable for long-term studies\footnote{\texttt{http://xmm2.esac.esa.int/docs/documents/\\CAL-TN-0018.pdf}}. 
Regarding the two available pn observations, we measured a $3\%$ relative increase in the blackbody temperature, which is formally significant within the typical errors of this instrument at the flux level of \jsixt. A similarly higher temperature is also measured in the most recent RGS1/2 observation with respect to the archival data. Contrarily to pn, such increase is not significant and is also not seen in the MOS data. The agreement of the derived spectral quantities is generally good between pn and RGS, and overall within the cross-calibration uncertainties between all instruments on-board \xmm\footnote{\texttt{ http://xmm.esac.esa.int/docs/documents/\\CAL-TN-0052.ps.gz}}. We therefore regard these discrepancies within the expected calibration uncertainties, rather than evidence of significant variability in the source physical properties within the time span of the analysis.

The results of our spectral fits are summarised in Table~\ref{tab_resultspec}. We list for each spectral model, numbered ($1-6$) as reference for the text, the reduced chi-squared $\chi^2_\nu$ and degrees of freedom (d.o.f.), the equivalent hydrogen column density $\nh$, model-dependent parameters (blackbody temperature $kT_\infty$, energy $\epsilon$ and width $\sigma$ of Gaussian absorption lines), line equivalent width ($EW$), and the unabsorbed source flux $F_{\rm X}$ in the $0.2-12$\,keV energy band (when considering simultaneous fits, we list the average between the detectors). The fit parameters were allowed to vary freely (unless otherwise noted), and within reasonable ranges -- in particular, restricting line energies between 0.3\,keV and 1.35\,keV, blackbody temperatures between $10$\,eV and $200$\,eV, and the Gaussian $\sigma$ between 5\,eV and 200\,eV. Uncertainties in the fit parameters represent $1\sigma$ confidence levels.
For the AO11 observation, results of spectral fitting in Table~\ref{tab_resultspec} are given for the pn camera alone and for the simultaneous fitting of the three EPIC instruments. 
Additionally, using the task \textsf{\small epicspeccombine}, we combined all 14 spectra and corresponding background and response files into one dataset; the results of spectral fitting of the combined spectrum\footnote{We note that these best-fit parameters are consistent with those resulting from the simultaneous fitting of all 14 datasets.} are also listed in Table~\ref{tab_resultspec}; folded models and residuals are shown in Fig.~\ref{fig_spec}. In the following, we discuss the spectral fitting in detail.
\begin{figure*}
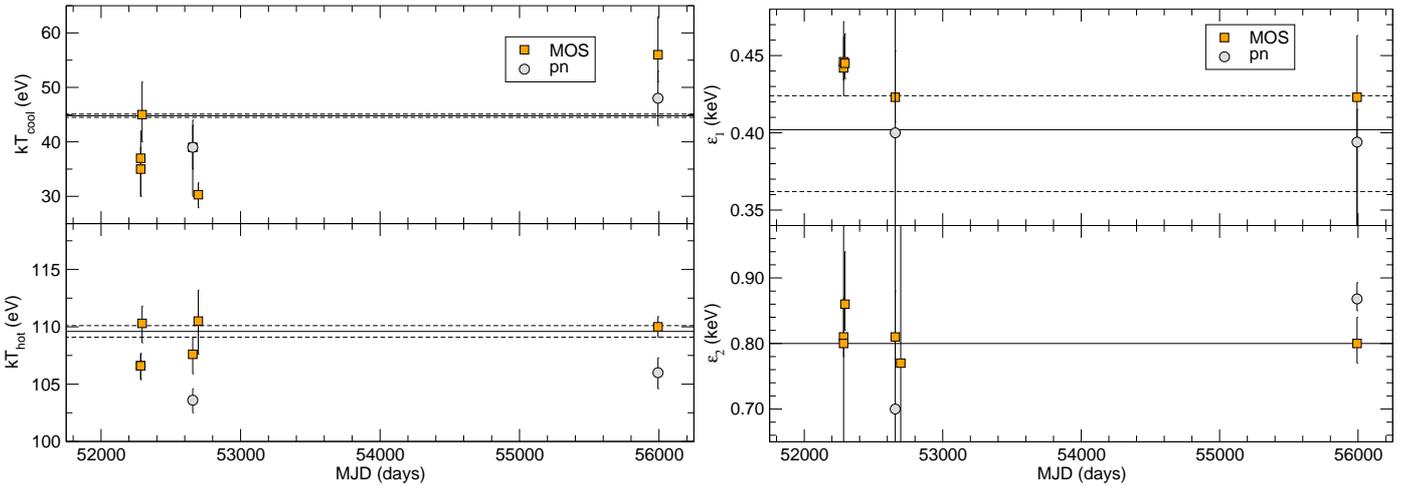

\begin{center}
\includegraphics*[width=0.495\textwidth]{fig5a.eps}
\includegraphics*[width=0.495\textwidth]{fig5b.eps}
\end{center}
\caption{Best-fit spectral parameters of \magonesix. Data points are results from fitting the MOS and pn spectra from the \xmm\ observations in Table~\ref{tab_pastxmmobs}. The spectral model corresponds to a double blackbody with two Gaussian absorption features (model (6) in Table~\ref{tab_resultspec}). We show the cool and hot best-fit blackbody components (left), as well as the best-fit line energies (right). Errors represent $1\sigma$ confidence level. Solid horizontal lines show results for the fit of the combined EPIC spectrum ($1\sigma$ errors are displayed as dashed lines).\label{fig_timebestparam}}
\end{figure*}

We first tried fitting a simple blackbody (1). The fit, as expected, is very unsatisfactory. The best-fit column density is very low and unconstrained. Absorption features at energies around $0.4$\,keV and $0.8$\,keV and an excess of counts at soft energies (which are seen in the residuals of all EPIC cameras) mostly contribute to the high value of $\chi^2$ (see Fig.~\ref{fig_spec}, top left).

The inclusion of a single Gaussian line in absorption at energy $\sim0.4$\,keV (2), while significantly improving the quality of the fit (lowering the reduced chi-square by a factor of 2 to 5, depending on the camera), is still a poor description of the spectrum of the source. The column density remains unconstrained and a somewhat lower temperature is found. Adding one more line in absorption to this simple blackbody model (3) lowers further the reduced chi-square, although not to a statistically acceptable value ($\chi^2_\nu=2.1$ for 86 degrees of freedom in the simultaneous EPIC fit). In general, the energy of the second line was found consistent with the ratio $\epsilon_2=2\epsilon_1$, which we discuss in Sect.~\ref{sec_disc2kT}.

As hinted by the excess of counts at soft energies, we found that the inclusion of a second cooler blackbody component, with best-fit temperature of $kT_{\rm cool}^\infty\sim35-40$\,eV, provides a much better description of the source continuum (4). However, the residuals at the line energies remain (see Fig.~\ref{fig_spec}, top right), and the column density tends to exceed the galactic value of $N_{\rm H}^{\rm gal}=2.4\times10^{20}$\,cm$^{-2}$ \citep{dic90} in this direction. 
Moreover, the emission radius as seen by an observer at infinity implied by the cool blackbody component is large, $R_{\rm cool}^\infty\sim30$\,km, assuming a distance of $d\sim350$\,pc \citep{pos07} to the source. 
Therefore, we restricted $\nh$ within 0 and $N_{\rm H}^{\rm gal}$ and added one (5) or two (6) Gaussian lines to the double blackbody continuum to look for acceptable fits.

With one line in absorption at a best-fit energy of $\epsilon_1\sim0.39$\,keV, a double blackbody model provides a fit with $\chi^2_\nu=1.6$ for 87 d.o.f., and a column density below the galactic value, $\nh=2.07(23)\times10^{20}$\,cm$^{-2}$ in the simultaneous EPIC fit. With respect to model (4), the temperature of the cool component is higher, $kT_{\rm cool}^\infty=70-75$\,eV, implying a radiation radius of $R_{\rm cool}^\infty\sim7$\,km. Significant residuals, especially seen in the pn (and in the combined) spectra, still remain at around the energy of the second feature, $\epsilon_2\sim0.85$\,keV (bottom-left plot of Fig.~\ref{fig_spec}), which motivates the inclusion of one further spectral component.

Finally, fitting the data with a double blackbody model and two Gaussian absorption lines results in a reduced chi-square of 1.5 for 85 d.o.f (in the simultaneous EPIC fit). The value of the best-fit column density is higher with respect to model (5), although still consistent within one standard deviation of $N_{\rm H}^{\rm gal}$. The temperature of the cool blackbody component, $kT_{\rm cool}=59.2_{-1.7}^{+1.5}$\,eV in the simultaneous EPIC fit, corresponds to a radiation radius of $R_{\rm cool}^\infty\sim11$\,km. The inclusion of a power-law tail extending towards higher energies, as usually seen in the emission of middle-aged pulsars dominated by soft thermal components (age $\sim$ few $10^5$\,yr, e.g. the ``Three Musketeers''; \citealt{luc05}), has no effect on the best-fit parameters of model (6) and does not significantly improve the $\chi^2$. We found that non-thermal power-law components, with typical photon indices of $\Gamma=1.7-2.1$ contribute at most $\sim0.3\%$ ($3\sigma$ confidence level, $0.2-12$\,keV range) of the unasorbed flux of the source. 
 
In general, consistent results were found between the detectors in the AO11 observation, as well as for the simultaneous fit of all spectra and observations. 
The fit results of spectral model (6) can be seen in Table~\ref{tab_resultspec} for each observation in Table~\ref{tab_pastxmmobs}. Where appropriate, a simultaneous fit of the two (MOS) or three (MOS+pn) EPIC detectors is performed for each observation. In Fig.~\ref{fig_timebestparam}, the temperature of the two blackbody components, as well as the energies of the two absorption lines, are plotted as a function of time (MJD). 
For comparison, horizontal lines show the corresponding values of fitting the combined EPIC spectrum, with $1\sigma$ confidence levels. Although significant variations are seen from observation to observation, they more likely reflect the still poor model description of the spectral energy distribution of the source, as well as to cross-calibration uncertainties between the detectors.
\subsubsection{RGS data\label{sec_rgsspec}}
Similarly to EPIC, we also included in the RGS spectral analysis all archival data not severely affected by background flares (Table~\ref{tab_pastxmmobs}). All observations were reduced in consistency with the AO11 data (Sect.~\ref{sec_rgsdatared}). 
To extract the source and background spectra, we used the standard instrument spatial masks and energy filters as well as GTI-filtered event lists. The source position, as inferred from the EPIC images (Table~\ref{tab_sourceMLparam}), was used to define the spatial extraction regions and the wavelength zero-point with the task \textsf{\small rgsregions}. Source and background spectra, and the corresponding redistribution matrix files in each RGS camera, were created with the tasks \textsf{\small rgsspectrum} and \textsf{\small rgsrmfgen}. Only the first order spectra are taken into account.
Due to the failure of chip~7 in RGS1 and of chip~4 in RGS2 (Sect.~\ref{sec_rgsdatared}), we ignored in the spectral fitting the corresponding defective channels of each instrument. As for the EPIC data, we allowed for a renormalisation factor between the two detectors; the energy band of the analysis is $0.35-1$\,keV.

The average flux in the RGS detectors, weighted by the errors ($f_{\rm X}^{\rm RGS}=4.37(27)\times10^{-12}$\,erg\,s$^{-1}$\,cm$^{-2}$ in energy band $0.35-2.5$\,keV), is consistent with a constant value ($\chi^2=5.4$ for 5 d.o.f.). The blackbody temperature between observations is also constant within errors (weighted mean $kT=97\pm3$\,eV; $\chi^2=3.1$ for 5 d.o.f.).
Since the source showed no evidence of significant variability in its spectral parameters, we co-added all spectra into two single files (one for each RGS detector), taking into account the different responses and background spectra, and binned the results to 0.18\AA\ to improve the signal-to-noise ratio. The total counts of the co-added RGS1/2 spectra in energy band $0.35-1$\,keV amount respectively to $5.8\times10^4$ and $5.1\times10^4$.

As for EPIC, the fit of a simple blackbody model, with best-fit $kT^\infty_{\rm hot}=96.4\pm0.5$\,eV, to the co-added RGS1/2 spectra clearly shows the excess of counts at soft energies. The inclusion of a second blackbody component, with best-fit $kT_{\rm cool}^\infty\sim25.0_{-1.9}^{+2.0}$\,eV, improves the quality of the fit by $\Delta\chi^2_\nu\sim1$. Due to the shorter spectral coverage of the RGS instrument at soft energies and the unconstrained $\nh$, the temperature of the cool blackbody component is significantly softer than that derived in the EPIC fits. Following the results from the EPIC analysis, we included two broad Gaussian features in absorption, with energies $\epsilon_1=443_{-20}^{+13}$\,eV ($\sigma_1=74_{-11}^{+14}$\,eV and equivalent width $31$\,eV) and $\epsilon_2=828(5)$\,eV ($\sigma_2=15\pm4$\,eV and equivalent width $13$\,eV). 
The inclusion of a narrow absorption feature as reported in \citet{ker04,hoh12a} gives a central energy of $\epsilon=576\pm{8}$\,eV and Gaussian width $\sigma=16^{+7}_{-5}$\,eV (equivalent width $5$\,eV), overall consistent with previous results. The column density, $\nh=2.5_{-0.8}^{+0.9}\times10^{20}$\,cm$^{-2}$, is consistent with the galactic value. 
The final reduced chi-square is $\chi^2_\nu=1.2$ for 213 d.o.f. 

\section{Discussion\label{sec_discussion}}
We summarise and discuss our findings in the light of the observed properties of the other members of the group of the ``Magnificent Seven'' isolated neutron stars. To help in the discussion we list in Table~\ref{tab_BPpropM7}, when appropriate for each of the seven objects, the temperature $kT_\infty$ of the best-fit single absorbed blackbody model; the spin period $P$ and pulsed fraction $p_{\rm f}$; the spin down $\dot{P}$, spin-down luminosity $\dot{E}=4.5\times10^{46}(\dot{P}P^{-3})$\,erg\,s$^{-1}$, dipolar magnetic field $B_{\rm dip}=3.2\times10^{19}(P\dot{P})^{1/2}$\,G, and characteristic timescale $\tau_{\rm ch}=P(2\dot{P})^{-1}$; the kinematic age from proper motion associations $t_{\rm kin}$; and the magnetic field derived from spectral proton cyclotron absorption \citep[e.g.][]{zan01}. For a recent compilation of the spectral and magneto-rotational properties of a sample of INSs with bright thermal X-ray emission, we refer to \citet{vig13}, as well as to these authors' on-line catalogue\footnote{\texttt{http://www.neutronstarcooling.info}}.

\begin{table*}[t]
\caption{Overall properties of the \msev
\label{tab_BPpropM7}}
\centering
\begin{tabular}{l c c c c c c c c c c}
\hline\hline
Object & $kT_\infty$ & $P$ & $p_{\rm f}$ & $\log(\dot{P})$ & $\log(\dot{E})$ & $\log(\tau_{\rm ch})$ & $\log(t_{\rm kin})$  & $\log(B_{\rm dip})$ & $\log(B_{\rm cyc})$ & Reference \\
       & (eV) & (s) & (\%)        & (s\,s$^{-1}$) & (erg\,s$^{-1}$) & (yr) & (yr) & ($10^{13}$\,G) & ($10^{13}$\,G) & \\ 
\hline
\magoneeig & $61$     & $7.06$ & 1  & $-13.527$ & $30.580$ & $6.58$ & $5.62$ & $13.17$ & $-$     & [1] \\
\magzersev & $84-93$  & $8.39$ & 11 & $-13.156$ & $30.726$ & $6.28$ & $5.93$ & $13.39$ & $13.75$ & [2] \\
\magonesix & $100$    & $3.39$ & 4  & $-11.796$ & $33.267$ & $4.53$ & $5.65$ & $13.87$ & $13.92$ & [3] \\
\magonethr & $100$    & $10.31$& 18 & $-12.951$ & $30.663$ & $6.16$ & $5.95$ & $13.54$ & $13.60$ & [4] \\
\magtwoone & $104$    & $9.43$ & 4  & $-13.398$ & $30.332$ & $6.57$ & $-$    & $13.29$ & $14.15$ & [5] \\
\magzereig & $95$     & $11.37$& 6  & $-13.260$ & $30.227$ & $6.51$ & $-$    & $13.40$ & $13.96$ & [6] \\
\magzerfou & $48$     & $3.45$ & 17 & $-13.553$ & $31.487$ & $6.29$ & $-$    & $13.00$ & $-$     & [7] \\
\hline
\end{tabular}
\tablefoot{The sources are sorted by decreasing brightness. 
\textit{References:} $^{[1]}$~\citet{kap02a,tie07,ker08,tet11,sar12,mig13}; $^{[2]}$~\citet{hab97,kap05a,tet11,hoh12b}; $^{[3]}$~this work, \citet{tet12}; $^{[4]}$~\citet{hab03,kap05b,schwope07,mot09}; $^{[5]}$~\citet{zan05,kap09a}; $^{[6]}$~\citet{hab02,kap09d}; $^{[7]}$~\citet{hab04b,kap11b}.}
\end{table*}
\subsection{Neutron star rotation\label{sec_discP}}
We have presented the results of a new \xmm\ observation of the \msev\ neutron star \magonesix. The higher count statistics and longer exposure (58\,ks) of the EPIC-pn camera with respect to past \xmm\ observations of the source have permitted the measurement of the period of the very likely neutron star rotation, at a value of $P\equiv P_\star=3.387864(16)$\,s that is the shortest amidst the \msev\ INSs, although comparable to that of \magzerfou\ (Table~\ref{tab_BPpropM7}). The pulsed fraction of the modulation is energy dependent and at best measured at 5\% for energies above $\sim0.5$\,keV, which correspond to a $\sim4\sigma$ detection in a blind search. We exclude the possibilty that the signal at $P_\star$ being associated with unknown instrumental effects, having found no significant signal in the $Z^2_n$ searches performed over photons extracted from several background regions, with roughly the same number of counts as collected for \jsixt\ (Sect.~\ref{sec_timinganalysis}). The analysis of the three EPIC cameras together more tightly constrains any periodicities with $p_{\rm f}\gtrsim1.6\%$ ($3\sigma$) and $P>5.2$\,s, thus considerably improving previous limits in this frequency range.

Our analysis shows that only the harder portion of the source spectrum -- namely, the photons with energy above $\sim0.5$\,keV, or roughly $\sim30\%$ of all source events -- shows a significant modulation within the sensitivity of our data. 
In the soft energy band of $0.2-0.5$\,keV, the pulsed fraction for pulsations in range $P=0.1468-100$\,s is constrained to be below $p_{\rm f}\lesssim3\%$ ($3\sigma$). Interestingly, \citet{tie07} also reported evidence for a pulsed fraction increasing as a function of energy in the EPIC-pn data of the brightest INS amongst the \msev, \magoneeig. Other members of the group similarly show changes in the pulse profile as a function of energy band, as well as phase-dependent spectral variations (of e.g. hardness ratio, depth of absorption lines; \citealt{hab03,hab04c}).

\begin{figure}
\begin{center}
\includegraphics*[width=0.5\textwidth]{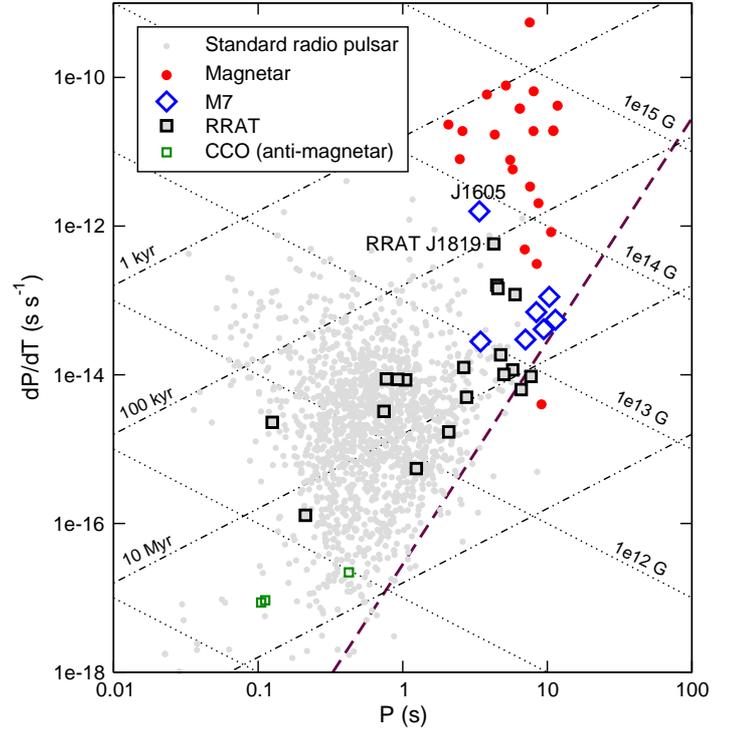}
\end{center}
\caption{$P-\dot{P}$ diagram highlighting the position of peculiar groups of isolated neutron stars (see legend). The tentative location of the \msev\ neutron star \magonesix\ is labelled, as that of the only rotating radio transient (RRAT) so far detected in X-rays, \rrat.\label{fig_PPdot}}
\end{figure}
\subsection{Candidate spin down and INS populations\label{sec_discPdotP}}
The analysis of the 2012 and 2003 EPIC-pn observations of \jsixt\ in a joined two-dimensional $Z^2_n(\nu,\dot{\nu})$ search (Sect.~\ref{sec_spindown}) provides a tentative candidate for the pulsar spin down of $\dot{\nu}=-1.4\times10^{-13}$\,Hz\,s$^{-1}$. Such spin down implies a dipolar magnetic field of $B_{\rm dip}\sim7.4\times10^{13}$\,G, under the usual assumption of magnetic braking in vacuum. A close estimate, $B_{\rm cyc}\sim8.3\times10^{13}$\,G, is derived from the line energy detected in the spectrum of the source, assuming that the feature at energy $\epsilon\sim0.4$\,keV is the fundamental proton cyclotron absorption (\citealt{zan01}; we further assume a canonical value of the gravitational redshift on the stellar surface of $z_g = 0.3$, for a neutron star mass and radius of $M_{\rm ns}=1.4$\,M$_\odot$ and $R_{\rm ns}=10$\,km, respectively). Despite the fact that our best spin-down solution being detected at a low confidence level ($2\sigma$), our $Z^2_n$ analysis excludes any other spin-down value at a significance higher than $1\sigma$, for $B_{\rm dip}=0-10^{14}$\,G. If confirmed, \jsixt\ would be the neutron star with the highest dipolar field amongst the \msev. 

In other members of the group, magnetic field estimates derived from both timing and spectral features similarly agree within a factor of a few, with the exception of the sources \magtwoone\ and \magzereig\ (see \citealt{kap09d,kap09a}, for a discussion; we note that the timing solutions of these INSs are not as well constrained as those of the other sources). In general, estimates derived from spectral lines are systematically higher than those from dipole braking. The absence of spectral lines in the spectrum of \magoneeig\ and \magzerfou\ also seems consistent with the weak magnetic field derived from spin down \citep{ker08,kap11b}.

We plot in Fig.~\ref{fig_PPdot} the tentative position of \jsixt\ in the $P-\dot{P}$ diagram of Galactic isolated neutron stars. The bulk of the population, characterised by the standard rotation-powered radio pulsars, is shown in the background whereas we highlight the position of peculiar groups of INSs, namely those of magnetars \citep[see e.g.][for a review]{mer08}, the \msev, rotating radio transients (RRATs, \citealt{lau06,kea11b}), and CCOs (a.k.a. anti-magnetars, \citealt{hal10,got10a}). 

The regions in the $P-\dot{P}$ diagram occupied both by the \msev\ and RRATs show considerable overlaps with those of the two opposite extremes of neutron star behaviour, observed namely in magnetars and ordinary radio pulsars. The discovery of RRATs, in particular, is intriguing since these sources have so far manifested themselves in a variety of ways. The most active source among the known sample is the highly magnetised \rrat, the only one\footnote{Of the nearly 70 objects known to date, $\sim20$ have precise position determinations and several have been investigated in X-rays, \citep{rea08,lau09a,kap09c,rea10}.} so far detected in X-rays \citep{lau07,cam13}. Very interestingly, the X-ray source was found to exhibit a spectrum remarkably similar to those of the \msev, although the RRAT is expected to be more distant and younger. Unusual timing behaviour following glitches detected in the radio indicates that this source could also have evolved from a magnetar \citep{lyn09}. It seems likely that several physical mechanisms are required to explain all the transient RRAT sources and the totality of distinct burst behaviours. This would imply that there is no unifying scenario that can account for all of their properties and that the intermitent radio emission might be due to factors both intrinsic and extrinsic to the neutron star as well as to selection biases. With the increasing number of newly discovered sources \citep[e.g.][]{den09b,kea11a}, these issues can be addressed and new scenarios can be proposed to explain their unusual radio emission.

On the other hand, the seven \ros-discovered INSs seem to form a rather homogenous class of cooling neutron stars. In particular, the sources show very similar timing and spectral behaviour, with perhaps only one exception: the long-term spectral variations of \magzersev\ \citep[][and references therein]{hoh12b}. 
Nonetheless, it is striking that a group of very similar sources, displaying at the same time unique properties that are so different from ordinary radio pulsars, are detected in the very local Solar vicinity. 
Indeed, the thermal X-ray luminosity of the \msev\ is in general higher than the power available from spin down, roughly by a factor of ten, which is in contrast to what is measured for radio pulsars detected at high energies, usually with $L_{\rm X}\lesssim10^{-1}\dot{E}$. Even higher factors of $L_{\rm X}/\dot{E}\sim10^2-10^3$ are measured for magnetars, suggesting that these neutron stars are too bright to be consistent with standard neutron star cooling \citep[e.g.][]{yak04}. 

The role played by the decay of the magnetic field in heating the neutron star crust, including state-of-art microphysics and the effect of Hall induction and Ohmic dissipation, has been investigated by \citet[][and references therein]{vig12,vig13}, in a worthy attempt to unify the several groups of INSs and explain the observed neutron star phenomenology. They found that, whereas for the bulk of the neutron star population the effect of the magnetic field on the luminosity is negligible, the magneto-thermal evolutionary models with original polar fields in range $B_{\rm p}^0\sim(3-5)\times10^{14}$\,G can account for the range of observed temperature, luminosity, age, as well as the timing properties of the \msev\ as the original field dissipates; magnetars with more extreme properties would require even higher original fields ($B^0_{\rm p}\gtrsim10^{15}$\,G) to explain their luminosities and timing properties. 

Our best $(P,\dot{P})$ solution implies a spin-down luminosity of $\dot{E}\sim1.8\times10^{33}$\,erg\,s$^{-1}$, which is significantly higher than the range measured for the other \msev\ stars, $\dot{E}\sim10^{30}-10^{31}$\,erg\,s$^{-1}$.
The spin period of $P\sim3.38$\,s is short for \msev\ standards, but comparable to that of the weakest-magnetised source \magzerfou. A fast-spinning neutron star would suggest the idea of a lower initial magnetic field and less dramatic spin down, in contrast to our result. However, the correlation between current magnetic field and period is not clear: both \magzereig\ and \magtwoone\ have long periods but relatively weak dipolar fields, to mention one example. 

The characteristic age derived from our best solution, $\tau_{\rm ch}\sim3.4\times10^4$\,yr, is also deviant from the typical few Myr measured for the other sources (Table~\ref{tab_BPpropM7}). Moreover, it implies a short evolutionary time scale that is inconsistent with the kinematic age derived from proper motion studies ($t_{\rm kin}=4.5\times10^5$\,yr; \citealt{tet12}). 
While spin-down ages are known to be reasonable estimators of a neutron star true age only under certain circumstances\footnote{Namely, if the initial rotation period was much shorter than the present value, and the magnetic field remained constant during the entire neutron star life; moreover, it is assumed that no additional braking due to e.g. the interaction with a fall-back disk has occurred, i.e. the pulsar is a perfect dipole rotator in vacuum.}, they usually overestimate the true pulsar age -- again, in contrast to our result. The (thus far) remarkably homogeneous timing properties of the \msev\ would argue against such a high value of spin down and discrepant associate quantities ($\dot{E}$, $\tau_{\rm ch}$). Additional data are required to definitely pin down the timing solution of this neutron star.
\subsection{Spectral energy distribution and lines in absorption\label{sec_disc2kT}}
The energy distribution of \jsixt\ in X-rays shows evidence for the presence of a cooler blackbody component with $kT_{\rm cool}^{\infty}\sim45-60$\,eV, as well as of two absorption features, in addition to the blackbody with $kT_{\rm hot}^\infty\sim100-110$\,eV. 

The detection of spectral features in absorption in several (non-accreting) thermally emitting neutron stars have been made possible thanks to the high sensitivity and spectral resolution of the \xmm\ and \chan\ observatories. Intriguingly, evidence for lines occurring at harmonically spaced energies has been suggested, where the best case study is the CCO \ccoonee\ (\citealt{san02,big03,mor05}; see e.g.\citealt{ker07a}, for the \msev). 
In contrast to accreting X-ray pulsars, where harmonics of the electron-cyclotron fundamental are essentially a relativistic phenomenon, \citet{sul10b} have shown that harmonically spaced features in CCOs can be associated to quantum effects in the energy dependence of the free-free opacity. 
Likewise in the \msev, where stronger magnetic fields of a few $10^{13}$\,G have been inferred from timing measurements, absorption lines can be associated with cyclotron processes of protons or ions. The occurrence (or suppression) of harmonics is, however, not clear, as a systematic investigation of such processes occurring in the atmosphere of these neutron stars has not yet (to our knowledge) been conducted so far.
Furthermore, the evidence for multiple lines in the observed spectrum of \jsixt\ has no unique interpretation, as they may consist of a blend of atomic (H, He) transitions in addition to a possible fundamental proton cyclotron line at $\sim0.4$\,keV.

High-resolution spectroscopy with the RGS instrument confirms the presence of a narrow absorption feature at energy $0.57$\,keV, possibly related with absorption of highly-ionised oxygen. \citet{hoh12a} investigated the RGS spectra of several bright INSs to find that narrow features are positively present in the spectra of those neutron stars that tend to be more distant and/or likely surrounded by a denser medium. The exact origin of the feature, if interstellar or in the atmosphere of the neutron star, is still under debate \citep[see e.g.][for a discussion]{hambaryan09,hoh12a}.

The distance estimated towards \jsixt\ of $d=350\pm50$\,pc -- derived from the amount of interstellar absorption in the X-ray spectrum of the source \citep{pos07} -- is not as reliably determined as in the cases where parallactic measurements are possible \citep{wal10,kap07}. It gives for our best model (Table~\ref{tab_resultspec}) emission radii of $R_{\rm cool}^\infty\sim11$\,km and $R_{\rm hot}^\infty\sim3$\,km. Non-thermal components extending towards higher X-ray energies are excluded at a level above $\sim0.3\%$ ($3\sigma$) of the source unabsorbed flux in the $0.2-12$\,keV energy band. The upper limit on the non-thermal luminosity is $L_{\rm X}^{\rm pl}(3\sigma)\sim3\times10^{29}$\,erg\,s$^{-1}$, much lower than the thermal X-ray component of a few $10^{32}$\,erg\,s$^{-1}$, or the rotational energy derived from our best $P,\dot{P}$ solution (Table~\ref{tab_BPpropM7}).
Evidence for a two-temperature model is also often found in the spectra of CCOs, some transient AXPs and that of the intriguing pulsar Calvera \citep{zan11}. In general, the emission radii (as seen by an a distant observer) are typically smaller than a few km, and are accompanied by unusual pulse profiles and strong pulsed fractions \citep[see][and the case of the CCO RX J0822-4300 in the supernova remnant Puppis~A]{got10b}, suggesting the framework of a model where hot spots are present at the neutron star surface. The large difference in intensity between the crustal and dipolar components of the magnetic field of the neutron star may indeed give origin to temperature anisotropies on the neutron star surface \citep{tur11,sha12}. 

The fact that most of the source photons show very low amplitude modulation, as well as the evidence for a double-temperature spectral model, might be used to constrain the temperature surface distribution and the geometry/orientation of the neutron star axes. Detailed phase-dependent spectral evolution studies, invoking an anisotropic temperature distribution based on polar caps with different temperatures and sizes, and not located at exact antipodal positions, have been carried out for several other neutron stars \citep[e.g.][]{schwope05,zan06a,got10b,ham11}. A similar approach for \jsixt, although not the scope of the present paper, is strongly encouraged.

\section{Summary and conclusions\label{sec_summary}}
Our most recent \xmm\ observation of the ``Magnificent Seven'' neutron star \magonesix\ has, most notably, revealed a candidate for its spin period at a value of $P\sim 3.38$\,s ($4\sigma$) that is comparable to the range observed in the other members of the group. Our analysis shows that the amplitude of the modulation is strongly energy dependent, and is only significantly detected at harder energies (in particular, roughly above the energy of the first spectral feature at energy $\sim0.5$\,keV). The coherent combination of the new data with a past \xmm\ EPIC-pn observation of the source constrains the pulsar spin-down rate at the $2\sigma$ confidence level, implying a dipolar magnetic field of $B_{\rm dip}\sim7.4\times10^{13}$\,G. If confirmed, this would rank the highest amongst the \msev. Spectral features in absorption, as well as a narrow absorption feature at energy $0.57$\,keV that is commonly observed in the spectra of other thermally emitting INSs, have also been positively identified. Moreover, we found compelling evidence for a two-temperature spectral distribution, which might explain the non-pulsating part of the source spectrum. Phase-resolved spectroscopy, as well as a dedicated observing campaign aimed at determining a timing solution, can give invaluable constraints on the neutron star geometry, allowing one to confirm a high value of spin down hinted by our analysis of the available \xmm\ data on the source.

\begin{acknowledgements}
The work of A.M.P. is supported by the Deutsche Forschungsgemeinschaft (grant PI~983/1-1). The authors acknowledge the use of the ATNF Pulsar Catalogue (\texttt{http://www.atnf.csiro.au/research/pulsar/psrcat}). We thank the anonymous referee for suggestions that helped improving the manuscript.
\end{acknowledgements}
\bibliographystyle{aa}
\bibliography{ins}
\end{document}